\documentclass[12pt]{article}
\usepackage{graphicx}
\usepackage{epsfig}
\usepackage{epstopdf}
\DeclareGraphicsExtensions{.pdf,.eps,.png,.jpg,.mps}
\usepackage{cite}

\def\lsim{\mathrel{\rlap{\lower3pt\hbox{\hskip0pt$\sim$}}
     \raise1pt\hbox{$<$}}}         
\def\gsim{\mathrel{\rlap{\lower4pt\hbox{\hskip1pt$\sim$}}
     \raise1pt\hbox{$>$}}}         

\usepackage{amsmath}
\usepackage{amsfonts}

\begin{document}
\begin{titlepage}

\centerline{\Large \bf Custom v. Standardized Risk Models}
\medskip

\centerline{Zura Kakushadze$^\S$$^\dag$\footnote{\, Zura Kakushadze, Ph.D., is the President of Quantigic$^\circledR$ Solutions LLC and a Full Professor at Free University of Tbilisi. Email: \tt zura@quantigic.com} and Jim Kyung-Soo Liew$^\sharp$\footnote{\, Jim Kyung-Soo Liew, Ph.D., is an Assistant Professor in Finance at the Johns Hopkins Carey Business School. Email: \tt kliew1@jhu.edu}}
\bigskip

\centerline{\em $^\S$ Quantigic$^\circledR$ Solutions LLC}
\centerline{\em 1127 High Ridge Road \#135, Stamford, CT 06905}
\centerline{\em $^\dag$ Free University of Tbilisi, Business School \& School of Physics}
\centerline{\em 240, David Agmashenebeli Alley, Tbilisi, 0159, Georgia}
\centerline{\em $^\sharp$ The Johns Hopkins Carey Business School}
\centerline{\em 100 International Drive, Baltimore, MD 21202}
\medskip
\centerline{(September 8, 2014; revised: May 7, 2015)}

\bigskip
\medskip

\begin{abstract}
{}We discuss when and why custom multi-factor risk models are warranted and give source code for computing some risk factors. Pension/mutual funds do not require customization but standardization. However, using standardized risk models in quant trading with much shorter holding horizons is suboptimal: 1) longer horizon risk factors (value, growth, {\em etc.}) increase noise trades and trading costs; 2) arbitrary risk factors can neutralize alpha; 3) ``standardized" industries are artificial and insufficiently granular; 4) normalization of style risk factors is lost for the trading universe; 5) diversifying risk models lowers P\&L correlations, reduces turnover and market impact, and increases capacity. We discuss various aspects of custom risk model building.
\end{abstract}

\bigskip
\medskip

\noindent{}{\bf Keywords:} Risk model, multi-factor, risk factor, short horizon, quant trading, style, industry, specific risk, factor risk, portfolio optimization

\end{titlepage}

\newpage

\section{Introduction}

{}In most incarnations, multi-factor risk models for stocks (RM) are based on style and industry risk factors. Industry factors are based on a similarity criterion, stocks' membership in industries under a given classification (e.g., GICS, ICB, BICS, etc.). Style factors are based on some estimated (or measured) properties of stocks. Examples of style factors are size \cite{Banz}, value and growth \cite{Basu, Fama.1992, Fama.1993, Lakonishok, AsnessStevens, Haugen, Liew}, momentum \cite{Jegadeesh, Asness}, liquidity \cite{Scholes, Pastor, AsnessKrail, Anson}, volatility \cite{Ang}, etc.\footnote{\, For an additional partial list (with some related literature), see, e.g., \cite{Black, Jensen, Blume, Brandt, Campbell.1987, Campbell.2001, Campbell.1988, Carhart, Chen, Cochrane, Connor, DeBondt, Fama.1996, Fama.1973, Ferson.1991, Ferson.1999, Hall, Jagannathan, Kothari, Lee, Lehmann, Lintner, Lo.2010, Lo.1990, MacKinlay, Merton, Mukherjee, Ng, Ross, Schwert, Sharpe, Whitelaw, Zhang}, and references therein. For a literature survey, see, e.g., \cite{HarveyLiu}.}

{}Most commercial RM are standardized (SRM). Majority of their users are institutions with longer term holdings (mutual funds, pension funds, etc.). To these users it is important that RM: i) contain fundamental longer horizon style risk factors (value, growth, etc.), and ii) be standardized, so their risk reporting and cross-institutional communications can be uniform. If a mutual fund risk report says ``Portfolio A has exposure B to the risk factor C under the risk model D'', its pension-fund client knows what that means. Typically, such portfolios are broad (low turnover allows lower cap/liquidity holdings) and fairly close to SRM coverage, so universe customization is not critical. Standardization is a much higher priority.

{}Quant trading with shorter-term holding portfolios has opposite priorities. Such strategies trade smaller universes of more liquid stocks (roughly, around 1,000-2,500 names), and these universes differ from strategy to strategy. Standardization is not critical for typically in-house risk reporting. Customization is much more important for a variety of reasons, which we discuss in more detail below. Here is a summary.

{}(1) {\em Short v. Long Horizons.} Strategies with shorter holding periods (e.g., statistical arbitrage and high frequency trading) do not benefit from longer horizon risk factors (value, growth, etc.) -- shorter-term returns are not highly correlated with quantities such as book value, which updates quarterly and lacks predictive power for holding horizons measuring in days or intraday. Moreover, when such risk factors are included in regression or optimization of shorter-term returns, they add noise to the holdings and trades, thereby increasing trading costs and reducing profitability.

{}(2) {\em Inadvertent Alpha Neutralization.} If, by design, some alphas are taking certain risk exposure, optimization using SRM typically will (partially) neutralize such alphas by producing a portfolio with more balanced risk exposure. E.g., if alphas are intentionally skewed toward small cap value, optimization using SRM, which typically contains ``size'' risk factor, will muddle such alphas by producing a portfolio with more balanced larger cap exposure. In such cases it is important to use custom RM (CRM) with the corresponding undesirable risk factors carefully omitted.

{}(3) {\em Insufficient Industry Granularity.} SRM coverage universe $U_{SRM}$ is ``squeezed'' into a modest number of fixed standardized ``industries'' (SRMI). Typically this reduces industry granularity, which adversely affects hedging industry risk. Thus, under a given industry classification (e.g., GICS, ICB, BICS, etc.), the true number of industries\footnote{\, Naming conventions vary by industry classification. ``Industry'' here refers to the most detailed level (i.e., the terminal branch) in a given classification tree (see Subsection \ref{ind.class} for details).} into which a given trading universe $U$ falls can be sizably higher than the number of SRMI. Also, different trading universes $U_1$ and $U_2$ fall into different (numbers of) true industries, while in SRM they are classified into the same SRMI.

{}(4) {\em Trading v. Coverage Universe.} Restricting $U_{SRM}$ to a substantially smaller trading universe $U$ yields side effects: i) for certain $U$ one can have empty SRMI (e.g., $U$ contains no telecom stocks), with no option to omit them from the factor covariance matrix (FCM) computation;\footnote{\, In fact, SRM sometimes may use a (small) subset $U_*$ of $U_{SRM}$ to compute FCM -- the required historical data may not be available for the entire $U_{SRM}$. However, it may be (and typically is) available for the trading universe $U$, so $U$, not artificial $U_*$, should be used for computing FCM.} and ii) it spoils normalization of style factors -- those normalized across $U_{SRM}$, e.g., conformed to a (log-)normal distribution.

{}(5) {\em Herd Effect.} Overusing SRM makes alphas more correlated, so when a shop blows up, it drags others with it (e.g., Aug'07 Quant Meltdown). This adversely affects shorter holding trading. Longer horizon strategies simply weather the storm.

{}(6) {\em Diversifying Risk Models Lowers P\&L Correlations.} When evaluating two RM against each other (in regression or optimization of the same returns), one looks at both A) the relative performance and B) P\&L correlation. If the P\&L correlation is low enough, it is more optimal to run a combined strategy using both RM thereby reducing portfolio turnover and market impact, and increasing capacity and P\&L.

{}(7) {\em Pros \& Cons of Custom Risk Models.} CRM further provide: (i) ability to compute RM based on custom universes and add/subtract risk factors; (ii) transparency; (iii) flexibility with update frequency (daily, weekly, etc.) and integration into test/production environments. However, the portfolio manager (PM) must compute CRM as opposed to receiving a file from an SRM provider.\footnote{\, Principal components provide ``customization'' to some extent. However, see footnote \ref{prin.comp}.} Data required to build CRM for shorter horizon trading is typically available to PM, albeit self-consistently computing FCM and idiosyncratic risk (ISR) is not common knowledge.

{}We discuss the above points (1)-(7) in more detail below. Section \ref{sec2} discusses RM in general, differences in their use in regression and optimization, and point (2). Section \ref{sec3} discusses decoupling of time horizons relevant to point (1). Section \ref{sec4} discusses points (3) and (4). We conclude in Section \ref{sec5} by discussing points (5)-(7). Appendix \ref{appA} contains R code for some style risk factors. Appendix \ref{appB} contains C code for symmetric matrix inversion. Appendix \ref{appC} contains legal disclaimers.

\section{Multi-factor Risk Models}\label{sec2}

{}In RM, a sample covariance matrix (SCM) $C_{ij}$ for $N$ stocks, $i,j = 1,\dots,N$  (computed based on time series of stock returns) is modeled by $\Gamma_{ij}$ given by
\begin{eqnarray}\label{Gamma}
 &&\Gamma \equiv \Xi + \Omega~\Phi~\Omega^T\\
 && \Xi_{ij} \equiv \xi_i^2 ~\delta_{ij}
\end{eqnarray}
where $\delta_{ij}$ is the Kronecker delta; $\Gamma_{ij}$ is an $N\times N$ matrix; $\xi_i$ is specific a.k.a. idiosyncratic risk (ISR) for each stock; $\Omega_{iA}$ is an $N\times K$ factor loadings matrix (FLM); and $\Phi_{AB}$ is a $K\times K$ factor covariance matrix (FCM), $A,B=1,\dots,K$. I.e., the random processes $\Upsilon_i$ corresponding to $N$ stock returns are modeled via $N$ random processes $\chi_i$ (ISR) together with $K$ random processes $f_A$ (factor risk):
\begin{eqnarray}
 &&\Upsilon_i = \chi_i + \sum_{A=1}^K \Omega_{iA}~f_A\\
 &&\left<\chi_i, \chi_j\right> = \Xi_{ij}\\
 &&\left<\chi_i, f_A\right> = 0\\
 &&\left<f_A, f_B\right> = \Phi_{AB}\\
 &&\left<\Upsilon_i, \Upsilon_j\right> = \Gamma_{ij}
\end{eqnarray}
The main reason for replacing the sample covariance matrix $C_{ij}$ by $\Gamma_{ij}$ is that the off-diagonal elements of $C_{ij}$ typically are not expected to be too stable out-of-sample. A constructed factor model covariance matrix $\Gamma_{ij}$ is expected to be much more stable as the number of risk factors, for which FCM $\Phi_{AB}$ needs to be computed, is $K\ll N$. Also, if $M<N$, where $M+1$ is the number of observations in each time series, then $C_{ij}$ is singular with $M$ nonzero eigenvalues. Assuming all $\xi_i>0$ and $\Phi_{AB}$ is positive-definite, then $\Gamma_{ij}$ is automatically positive-definite (and invertible).

\subsection{Industry Risk Factors}\label{ind.class}

{}RM can mix risk factors of different types. Typically, the most numerous factors are industry factors. In its simplest incarnation, industry FLM $\Omega^{\rm{\scriptstyle{ind}}}_{iA}$, $A=1,\dots,K_{\rm{\scriptstyle{ind}}}$ is a binary matrix of 1s and 0s indicating whether a stock belongs to a given industry:
\begin{eqnarray}
 &&\Omega^{\rm{\scriptstyle{ind}}}_{iA} = \delta_{G(i), A}\\
 &&G:\{1,\dots,N\} \mapsto \{1,\dots,K_{\rm{\scriptstyle{ind}}}\}\\
 &&\sum_{i=1}^N \Omega^{\rm{\scriptstyle{ind}}}_{iA} = N_A\\
 &&\sum_{A=1}^{K_{\rm{\scriptstyle{ind}}}} \Omega^{\rm{\scriptstyle{ind}}}_{iA} = 1
\end{eqnarray}
where $G$ is the map between stocks and industries, and $N_A > 0$ is the number of stocks in the industry labeled by $A$ (we are assuming there are no empty industries).

{}More generally, we can allow each stock to belong to multiple industries (e.g., in the case of conglomerates) with some weights $\omega_{iA}$:
\begin{eqnarray}
 &&\Omega^{\rm{\scriptstyle{ind}}}_{iA} = \sum_{B\in G(i)}\omega_{iB}~\delta_{AB}\\
 &&\sum_{B\in G(i)}\omega_{iB} = 1
\end{eqnarray}
where $G(i)\subset \{1,\dots,K_{\rm{\scriptstyle{ind}}}\}$ and $\left|G(i)\right| \equiv n_i$ need not be 1, albeit typically for most stocks $n_i = 1$, and (for conglomerates) $n_i\neq 1$ is not a large number.

{}Binary $\Omega^{\rm{\scriptstyle{ind}}}_{iA}$ can be constructed from binary classifications such as GICS, ICB, BICS, etc. Naming conventions for levels differ by classification, e.g., sector, sub-sector, industry, sub-industry, etc. ``Industry'' here means the most detailed level in the classification tree. If there are too many industries (e.g., with small $N_A$), one can ``prune'' the tree by merging (small) industries at higher levels, e.g., in the tree ``sector $\rightarrow$ sub-sector $\rightarrow$ industry'', industries are pruned at the sub-sector level.

\subsection{Style Risk Factors}

{}Industry factors are based on a similarity criterion: industry membership. Style risk factors are based on some estimated (or measured) properties of stocks. Examples of style factors are size, liquidity, volatility, momentum, growth, value, etc.

{}For illustrative purposes, let us discuss some style factors in more detail. In Appendix \ref{appA} we give R code for size, liquidity, intraday volatility and momentum style factors.\footnote{\, Legal disclaimers regarding this code are included in Appendix \ref{appC}.}

{}$\bullet$ Size is a logarithm of market cap, normalized (conformed to normal distribution) -- see R code in Appendix \ref{appA}. Note that ADRs are treated separately.

{}$\bullet$ Liquidity is a logarithm of the average daily dollar volume (ADDV),\footnote{\, E.g., over the past 20 trading days. One may prefer to take, say, last 3 months.} normalized similarly to size with ADRs treated separately -- see Appendix \ref{appA}.

{}$\bullet$ Volatility style factor can be based on historical (relevant for longer-term models) or intraday (relevant for shorter-term models -- see the next section) volatility. One way to define intraday volatility is to use intraday high and low prices -- see Appendix \ref{appA}. The log of intraday volatility is normalized similarly to size.

{}$\bullet$ Momentum can be defined as a normalized average over $D$ trading days of $d$-trading-day moving-average returns (e.g., $d=5$, $D=252$)\footnote{\, In the 0-th approximation, this is a $D$-day return. Removing outliers introduces $d$-dependence.} -- see Appendix \ref{appA}.

{}$\bullet$ Value can be defined via a normalized book-to-price ratio (negative values need to be dealt with). Growth requires earnings data. As discussed below, longer horizon style factors such as value and growth do not add value in shorter horizon strategies as book value and earnings data essentially are updated quarterly.

\subsection{Factor Covariance Matrix and Specific Risk}\label{fac.cov}

{}Once FLM $\Omega_{iA}$ is defined,\footnote{\, One can also use the first $K_{\rm{\scriptstyle{prin}}}$ principal components (PC) of SCM $C_{ij}$ as columns in FLM $\Omega_{iA}$. However, the out-of-sample instability in the off-diagonal elements of SCM is also inherited by PC. Furthermore, if $M < N$, SCM is singular and only $M$ PC are available. It is for these reasons that style and industry factors are more widely used in practical applications.\label{prin.comp}} FCM $\Phi_{AB}$ and ISR $\xi_i$ must be constructed. Straightforwardly computing FCM as a sample\footnote{\, Out-of-sample stability and singularity of FCM when $M < K$ are issues to consider.} covariance matrix of the risk factors $f_A$ is insufficient as ISR $\xi_i$ computed using such sample FCM typically are ill-defined. Algorithms for consistently computing FCM and ISR usually are deemed proprietary.

\subsection{Use of Factor Models}

{}RM have a variety of uses, which are not limited to passively measuring risk exposures, but also include actively hedging such risks, e.g., by requiring that a portfolio be neutral w.r.t. various risk factors and/or such risk exposures be optimized.\footnote{\, See, {\em e.g.,} \cite{Kakushadze} for a recent discussion.} Here we outline the usage of RM in regression and optimization.

\subsubsection{Regression}

{}Let $R_i$, $i=1,\dots,N$ be the stock expected returns. A weighted regression (without intercept) of $R$ over FLM $\Omega$ with weights $z_i$ is given by (in matrix notation):\footnote{\, This is a cross-sectional regression; in R notations $\varepsilon = \mbox{residuals}(\mbox{lm}(R \sim \Omega - 1, \mbox{weights} = Z))$.}
\begin{eqnarray}\label{reg.res}
 &&\varepsilon \equiv R - \Omega~Q^{-1}~\Omega^T~Z~R\\
 \label{w.reg.2}
 &&Z \equiv\mbox{diag}(z_i)\\
 \label{w.reg.3}
 &&Q\equiv \Omega^T~Z~\Omega\\
 \label{w.reg.4}
 &&{\widetilde R} \equiv Z~\varepsilon
\end{eqnarray}
Here $\varepsilon_i$ are the residuals of the weighted regression. Also, note that the ``regressed'' returns ${\widetilde R}_i$ are neutral w.r.t. the $K$ risk factors corresponding to $\Omega_{iA}$:
\begin{equation}\label{neut}
 \sum_{i=1}^N{\widetilde R}_i~\Omega_{iA} = 0,~~~A=1,\dots,K
\end{equation}
If $\Omega_{iA}$ includes the intercept (a unit vector) as one of its columns, then we have
\begin{equation}\label{d.n}
 \sum_{i=1}^N {\widetilde R}_i = 0
\end{equation}
I.e., in this case the regressed returns are demeaned. The weights $z_i$ can be chosen to be unit weights, or, e.g., $z_i\equiv 1/\sigma_i^2$, where $\sigma_i$ is some historical volatility of $R_i$.

{}Using ${\widetilde R}_i$, a simple mean-reversion strategy can be constructed via
\begin{equation}\label{m.r}
 D_i = -\gamma~{\widetilde R}_i
\end{equation}
where $D_i$ are the desired dollar holdings and $\gamma > 0$ is fixed via
\begin{equation}\label{inv.lvl}
 \sum_{i=1}^N |D_i| = I
\end{equation}
where $I$ is the total desired investment level. The portfolio (\ref{m.r}) is dollar neutral if we have (\ref{d.n}). For this weighted regression all we need is FLM $\Omega_{iA}$ -- FCM $\Phi_{AB}$ is not needed,\footnote{\, Rotating $\Omega_{iA}$ by an arbitrary $K\times K$ nonsingular matrix $U_{AB}$, $\Omega\rightarrow \Omega ~U$, does not change the regression residuals (\ref{reg.res}) or the risk neutrality conditions (\ref{neut}).} nor is ISR $\xi_i$, albeit if the latter are known, they can be used (instead of $\sigma_i$) in the regression weights via $z_i\equiv 1/\xi_i^2$ (see \cite{Kakushadze} for details).

\subsubsection{Optimization}

{}In optimization, rather than requiring strict neutrality w.r.t. the risk factors, one can require that risk exposure be optimized, albeit one can do both (see below). In its simplest incarnation, optimization requires that the Sharpe ratio of the portfolio (using the notations of the previous subsection)
\begin{equation}
 S = {{\sum_{i=1}^N D_i R_i} \over\sqrt{\sum_{i,j=1}^N C_{ij} D_i D_j}} \rightarrow \mbox{max}
\end{equation}
Assuming no costs, constraints or bounds, the Sharpe ratio is maximized by
\begin{equation}
 D_i = \zeta\sum_{j=1}^N C^{-1}_{ij} R_j
\end{equation}
where $\zeta$ is fixed via (\ref{inv.lvl}). In the RM context, one replaces $C_{ij}$ via $\Gamma_{ij}$, which gives:\footnote{\, This follows from the expression for the inverse of $\Gamma$: $\Gamma^{-1} = \Xi^{-1} - \Xi^{-1}~\Omega~{\widetilde Q}^{-1}~\Omega^T~\Xi^{-1}$.}
\begin{eqnarray}\label{d.h}
 &&D_i = {\zeta\over \xi^2_i}~\left(R_i - \sum_{j = 1}^N {R_j \over \xi^2_j}~\sum_{A,B = 1}^K \Omega_{iA}~\Omega_{jB}~{\widetilde Q}^{-1}_{AB} \right)\\
 &&{\widetilde Q}_{AB}\equiv \Phi^{-1}_{AB} + \sum_{i = 1}^N {1\over \xi^2_i}~\Omega_{i A}~\Omega_{i B}\label{Q.tilde}
\end{eqnarray}
where $\Phi^{-1}_{AB}$ is the inverse\footnote{\, Appendix \ref{appB} contains C code for symmetric matrix inversion.} of $\Phi_{AB}$, and ${\widetilde Q}^{-1}_{AB}$ is the inverse of ${\widetilde Q}_{AB}$. The desired dollar holdings $D_i$ are not neutral w.r.t. the risk factors, nor are they dollar neutral. Dollar and/or various risk factor neutrality can be achieved via optimization with homogeneous linear constraints (see \cite{Kakushadze}). Note that to compute the desired dollar holdings (\ref{d.h}), we need not only FLM $\Omega_{iA}$, but also FCM $\Phi_{AB}$ and ISR $\xi_i$. This is a key difference between using RM in optimization {\em vs.} regression.

\subsubsection{``Risk-taking'' Alphas}\label{risk.taking}

{}While in optimization the resulting desired dollar holdings $D_i$ are not neutral w.r.t. the risk factors, they are ``approximately'' neutral in the sense that the deviation from neutrality is due to ISR. Indeed, from (\ref{d.h}) we have
\begin{eqnarray}\label{r.n.1}
 &&\sum_{i=1}^N D_i~\Omega_{iA} = \zeta \sum_{i=1}^N {R_i \over \xi^2_i}~\Omega_{iB}~\Delta^{-1}_{BA}\\
 &&\Delta_{AB} \equiv \delta_{AB} + \sum_{i=1}^N \sum_{C=1}^K {1\over\xi_i^2}~\Phi_{AC}~\Omega_{iC}~\Omega_{jB}\label{r.n.2}
\end{eqnarray}
where $\Delta^{-1}$ is the inverse of $\Delta = \Phi~{\widetilde Q}$. Let $\Phi_{AB}\equiv\kappa~{\widehat\Phi}_{AB}$, and let us take the limit $\kappa\rightarrow \infty$ with ${\widehat\Phi}_{AB} = \mbox{fixed}$. In this limit the factor risk dominates and ISR is negligible, so optimization reduces to the weighted regression of the previous sub-subsection (see \cite{Kakushadze} for details), and we have
\begin{equation}
 \sum_{i=1}^N D_i~\Omega_{iA}\rightarrow 0
\end{equation}

{}So, regression yields risk neutrality, while optimization produces approximate risk neutrality. Either way, if the returns $R_i$ have exposure to a risk factor, it is either eliminated (regression) or substantially reduced (optimization). This has important implications for using SRM in certain types of trading.

{}Thus, imagine that {\em by design} the returns $R_i$ have {\em desirable} risk exposure, e.g., our strategy could be deliberately skewed toward small cap value stocks, have exposure to momentum, volatility, etc.\footnote{\, Real-life alphas often have sizable exposure to risk -- a real-life alpha is any reasonable expected return. E.g., momentum strategies often have substantial exposure to risk. Furthermore, there is no ``perfect'' risk model. Otherwise, there would only be mean-reversion caused by temporary trading imbalances. For a complementary discussion, see, e.g., \cite{Lee}.} If such a risk factor is included in FLM $\Omega_{iA}$, then using the full FLM in regression and/or optimization would be to the detriment of our alpha. In the regression we can simply omit the corresponding risk factor(s). However, in the optimization merely omitting risk factors will not do -- we also need to recompute FCM and ISR anew based on the remaining risk factors, otherwise we will get wrong predictions for the total risk:
\begin{equation}
 \Gamma_{ii}^\prime \equiv \xi_i^2 + \sum_{A^\prime,B^\prime\in H} \Omega_{iA^\prime}\Omega_{iB^\prime}~\Phi_{A^\prime B^\prime} \neq \Gamma_{ii}
\end{equation}
where $H\subset\{1,\dots,K\}$ is the subset corresponding to the remaining risk factors. In this case SRM simply will not do and CRM is required.

\section{Decoupling of Time Horizons (Frequencies)}\label{sec3}

{}There is an important fundamental concept, which can be stated as decoupling of time horizons (or, equivalently, frequencies or wavelengths). In a nutshell, what happens at time horizon $T_1$ (or frequency $f_1 = 1/T_1$) is not affected by what happens at time horizon $T_2$ (or frequency $f_2 =1/T_2$) if $T_1$ and $T_2$ are vastly different. By time horizon we mean the relevant time scales. E.g., the time horizon for a daily close-to-close return is 1 day. In terms of returns, the decoupling can be restated as the returns for long-term horizons $T_1$ being essentially uncorrelated with the returns for short-term horizons $T_2$. Here is a simple argument.

\subsection{Short v. Long Time Horizons}

{}Here is a simple argument for a single stock (or security). Consider a time interval from time $t_0$ to time $t_M > t_0$. Let us divide it into $M$ intervals $t_0, t_1,\dots,t_{M-1},t_M$. For simplicity, we can assume that these intervals are uniform, $t_s = t_0 + s ~\Delta t$, $s = 0,\dots,M$, albeit this is not critical here. Let the stock prices at times $t=t_s$ be $P(t_s)$. Let us define the return from time $t$ to time $t^\prime$ as
\begin{equation}
 R(t,t^\prime) \equiv \ln\left({P(t^\prime)\over P(t)}\right)
\end{equation}
Then we have
\begin{eqnarray}
 &&{\widetilde R} \equiv R(t_0,t_M) = \sum_{s=1}^M R_{s}\\
 &&R_{s}\equiv R(t_{s-1},t_s)
\end{eqnarray}

{}Now let us ask the following question. How correlated is the return $R_{M}$ for the most recent period (i.e., $t_{M-1}$ to $t_M$) with the return ${\widetilde R}$ for the entire period (i.e., $t_0$ to $t_M$)? To define a ``correlation'', we need multiple observations. So, we hang yet another index onto our returns, call it $\alpha$, where $\alpha = 1,\dots,p$ labels different periods $t_0^\alpha$ to $t_M^\alpha$, each consisting of $M$ periods ($t^\alpha_{s-1}$ to $t^\alpha_s$, $s=1,\dots,M$), and we wish to compute the correlation between the return for the last such period $R_M^\alpha$ and the return for the entire such period ${\widetilde R}^\alpha$, and $\alpha$ (not $s$) labels the series (which is a time series) over which the correlation is computed. For simplicity, we can assume that the periods labeled by $\alpha$ are ``tightly packed'', i.e., $t^{\alpha-1}_{M} = t^{\alpha}_{M-1}$, albeit this is not crucial here. We then have $p+M$ time points $\tau_r$, $r = 0,1,\dots, p+M-1$ and, consequently, $p+M-1$ returns ${\widehat R}_{r}$, $r=1,\dots, p+M-1$, where
\begin{eqnarray}
 &&\tau_r \equiv t_0 + r~\Delta t,~~~r = 0,1,\dots, p+M-1\\
 &&{\widehat R}_{r} \equiv \ln\left({P(\tau_r)\over P(\tau_{r-1})}\right),~~~r = 1,\dots, p+M-1\\
 &&R^\alpha_{s} = {\widehat R}_{s + \alpha - 1}, ~~~s=1,\dots,M,~~~\alpha = 1,\dots,p\\
 &&{\widetilde R}^\alpha = \sum_{r=\alpha}^{M+\alpha-1}{\widehat R}_r
\end{eqnarray}
Note that $p$ can be much larger than $M$, in fact, we will assume this to be the case.\footnote{\, To illustrate, if, say, $\Delta t$ is 1 day, then we are computing the correlation of the $M$-day moving average return ${\overline R}^\alpha \equiv {1\over M} {\widetilde R}^\alpha$ with the last daily return in the moving average, and we have $p$ rolling periods like this. We have $p+M$ dates and, consequently, $p+M-1$ daily returns.}

{}With the covariance $\langle *,*\rangle$ defined as above, we have
\begin{equation}
 \sigma_{s}^2 \equiv \langle R_{s}, R_{s}\rangle = {1\over p} \sum_{\alpha = 1}^p {\widehat R}^2_{s + \alpha - 1} - {1\over p^2} \left(\sum_{\alpha = 1}^p {\widehat R}_{s + \alpha - 1}\right)^2\approx \sigma^2
\end{equation}
where
\begin{equation}
 \sigma^2 \equiv {1\over p} \sum_{r = 1}^p {\widehat R}^2_{r} - {1\over p^2} \left(\sum_{r = 1}^p {\widehat R}_{r}\right)^2
\end{equation}
and we have used the fact that $p\gg M$, which implies that all $M$ variances $\sigma_{s}^2$ are approximately the same. We then have
\begin{eqnarray}
 &&\langle R_{s}, R_{s^\prime}\rangle \equiv \sigma_{s}~\sigma_{s^\prime}~\Psi_{ss^\prime}\approx \sigma^2 ~\Psi_{ss^\prime}\\
 &&\langle {\widetilde R}, {\widetilde R}\rangle = \sum_{s,s^\prime = 1}^M \langle R_{s}, R_{s^\prime}\rangle \approx \sigma^2 \sum_{s,s^\prime = 1}^M\Psi_{ss^\prime}\\
 &&\langle R_{s}, {\widetilde R}\rangle = \sum_{s^\prime=1}^M \langle R_{s}, R_{s^\prime}\rangle \approx \sigma^2 \sum_{s^\prime = 1}^M\Psi_{ss^\prime}
\end{eqnarray}
where $\Psi_{ss^\prime}$ is the $M\times M$ correlation matrix of returns $R_{s}$, and $\Psi_{ss} = 1$. We have
\begin{equation}\label{chi.s}
 \rho_{s}\equiv \mbox{Cor}(R_{s}, {\widetilde R})\approx {\sum_{s^\prime = 1}^M\Psi_{ss^\prime}\over\sqrt{\sum_{s,s^\prime = 1}^M\Psi_{ss^\prime}}} =
 {{1 + \psi_{s}}\over\sqrt{M + \sum_{s=1}^M \psi_{s}}}
\end{equation}
where
\begin{equation}
 \psi_{s} \equiv \sum_{s^\prime\in J_s} \Psi_{ss^\prime}
\end{equation}
and $J_s \equiv\{1,\dots,M\}\setminus\{s\}$.

{}Because we have $M\ll p$, the matrix $\Psi_{ss^\prime}$ is approximately ``self-similar'' in the sense that all $m \times m$ sub-matrices of $\Psi_{ss^\prime}$ with $s,s^\prime \in\{k,k+1,\dots,k+m-1\}$, $k=1,\dots,M-m+1$ and $1 < m < M$ (i.e., for each $m$ there are $M-m+1$ such sub-matrices) are approximately the same. Put differently, $\Psi_{ss^\prime}$ approximately depend only on the difference $s-s^\prime$, and, in fact, only on $|s-s^\prime|$ since $\Psi_{ss^\prime}$ is symmetric. Let $\Psi_{s,1} \equiv \eta_{s-1}$, $s>1$. Then we have $\Psi_{ss^\prime} \approx \eta_{|s - s^\prime|}$, $s\neq s^\prime$, and
\begin{eqnarray}
 &&\psi_M \approx \sum_{s=1}^{M-1}\eta_s\\
 &&\sum_{s=1}^M \psi_{s}\approx 2\sum_{s=1}^{M-1}(M-s)~\eta_s
\end{eqnarray}
To estimate the correlation $\rho_M$, we need to make some assumptions about the correlations $\eta_s$. A reasonable assumption is that the correlations $\Psi_{ss^\prime}$ decay as $|s-s^\prime|$ grows. E.g., we can assume that $|\Psi_{ss^\prime}| \leq \lambda^{|s-s^\prime|}$ for some positive $\lambda < 1$, or, equivalently, that $\eta_s = {\widetilde \eta}_s~\lambda^s$, $s=1,\dots,M-1$, where $|{\widetilde\eta}_s|\leq 1$. We then have
\begin{eqnarray}
 &&\psi_M\approx f(\lambda)\equiv \sum_{s=1}^{M-1}{\widetilde \eta}_s~\lambda^s\\
 &&\sum_{s=1}^M \psi_s \approx  2\left[M~f(\lambda) - \lambda~f^\prime(\lambda)\right]\\
 &&f^\prime(\lambda)\equiv {\partial f(\lambda)\over\partial\lambda}
\end{eqnarray}
and our correlation $\rho_M$ reads
\begin{equation}\label{chi.M}
 \rho_M\approx {{1+f(\lambda)}\over\sqrt{M\left[1 + 2~f(\lambda)] - 2\lambda~f^\prime(\lambda)\right]}}
\end{equation}
We also have
\begin{eqnarray}
 &&|f(\lambda)| \leq {{\lambda-\lambda^M}\over{1-\lambda}}\approx {\lambda\over {1-\lambda}}\\
 &&|f^\prime(\lambda)| \leq {{1 + (M-1)~\lambda^M - M~\lambda^{M-1}}\over(1-\lambda)^2}\approx {1\over(1-\lambda)^2}
\end{eqnarray}
where we have taken into account that $M\gg 1$.

{}So, if $f(\lambda)\geq 0$, we have (assuming $M$ is large)
\begin{equation}\label{chi_M}
 0<\rho_{M}\lsim {1\over(1-\lambda)\sqrt{M}} \ll 1
\end{equation}
and for positive $f(\lambda)$ the bound is even tighter.

{}What about $f(\lambda) < 0$? First, note that $\rho_M$ is still positive -- for it to become negative, the argument of the square root in the denominator in (\ref{chi.M}) would become negative, which is not possible (see below). However, for negative $f(\lambda)$ {\em a priori} it might appear that $\rho_M$ need not be small as the denominator in (\ref{chi.M}) could become small when $f(\lambda) = -1/2 + \epsilon$, where $0 < \epsilon \sim 1/M$. Nonetheless, this cannot be the case if there is randomness in the returns $R_s$. Indeed, we have\footnote{\, Note that $\rho_M > 0$ unless $f(\lambda) \leq -1$, for which $ \langle {\widetilde R}, {\widetilde R}\rangle$ would be negative considering $M \gg 1$.}
\begin{equation}
 \langle {\widetilde R}, {\widetilde R}\rangle \approx \sigma^2 \sum_{s,s^\prime = 1}^M\Psi_{ss^\prime} \approx \sigma^2\left(M\left[1 + 2~f(\lambda)] - 2\lambda~f^\prime(\lambda)\right]\right)
\end{equation}
So, the argument of the square root in the denominator of (\ref{chi.M}) is (up to $\sigma^2$) the variance of the return ${\widetilde R}$ for the period $t_M-t_0 = M~\Delta t$. If there is randomness in the returns $R_s$, the variance $\langle {\widetilde R}, {\widetilde R}\rangle$ should scale linearly with $t_M-t_0$ and, consequently, with $M$. If this variance were of order $\sigma^2$, this would imply that the returns $R_s$ are highly anti-correlated with each other and the entire process is highly deterministic. Put differently, there would be essentially no dispersion in this case. Under normal circumstances, where we have randomness in the returns $R_s$, the variance $\langle {\widetilde R}, {\widetilde R}\rangle$ should be of order $M~\sigma^2$. If there are any negative correlations $\eta_s$, they are offset by other positive correlations so that $\langle {\widetilde R}, {\widetilde R}\rangle \sim M~\sigma^2$ and we have (\ref{chi_M}).

{}The upshot is -- this is a generalization of our example above -- that quantities with long time horizons have low correlations with quantities with short horizons. What happens, say, at milliseconds gets diluted by the time one gets to, say, month-long horizons -- and this dilution is due to the cumulative effect of everything that transpires in between such vastly different time scales. Randomness plays a crucial role in this dilution. If things were deterministic, such dilution would not occur.\footnote{\, This is ``analogous'' to what happens in quantum mechanics and quantum field theory. We put the adjective ``analogous'' in quotation marks because a stochastic process described by Brownian motion is nothing but Euclidean quantum mechanical particle, so the ``analogy'' is in fact precise.}

\subsection{Implication for Risk Factors}\label{sub3.1}

{}A practical implication of the above discussion is that care is needed in choosing which risk factors to use in RM depending on what the time horizons of the strategies are for which RM is used. If these horizons are short, then risk factors such as value and growth, whose underlying fundamental data updates quarterly, should not be used as they add no value in short holding (a few days, overnight, intraday, etc.) strategies. Here is a simple argument. Consider high frequency trading at, say, millisecond time scales. Does book value make a difference to such trading? The answer is no. What is relevant here is the market microstructure at the millisecond timescales (bid, ask, bid and ask sizes, order book depth, hidden liquidity, posting orders fast on different exchanges, whether the trader's collocation is close to the exchange connectivity hub, etc.).\footnote{\, Similarly, growth does not add value in this context either. This is not to say that, e.g., earnings are not important in short-term trading. However, the way to implement them is via monitoring earnings announcements and, e.g., not trading stocks immediately following their earnings announcements, not by using growth style factor in, say, intraday regressions or optimization.} Whether the book value for stock XYZ is \$100M or \$1B does not directly affect the market microstructure at millisecond time scales.\footnote{\, Arguably, there might be higher-order indirect effects via the book value affecting liquidity and market cap (see below). However, such higher-order effects are expected to be lost in all the noise. They might be ephemerally amplified around the time book value is updated (quarterly).}

{}On the other hand, quantities such as liquidity and market cap\footnote{\, Market cap is relevant primarily because it is highly correlated with liquidity.} do affect market microstructure. E.g., liquidity affects typical bid/ask sizes, print sizes, etc. More precisely, liquidity computed based on, say, 20-trading-day ADDV indirectly relates to such ``micro'' quantities because of the expected {\em linear scaling} of volumes.\footnote{\, One can directly measure intraday liquidity based on ``micro'' quantities, which is more tedious. Typically, ADDV based computation reasonably agrees with such ``micro'' computation.} I.e., even though ADDV is computed using longer horizons, it is a relevant risk factor for shorter horizon strategies precisely because of the aforementioned linear scaling of volumes, allowing an {\em extrapolation} from longer to shorter horizons.

{}Similarly, volatility is a relevant style factor. Typically, it is computed as historical volatility of, say, close-to-close returns. As an {\em extrapolation} -- based on the {\em assumption} that historically more volatile stocks are also more volatile intraday -- one can use this style factor for shorter horizon strategies. Preferably, one can also define volatility style factor based on shorter horizons (e.g., intraday; see Section \ref{sec2}).

{}So, conceptually, if the underlying quantity (e.g., book value or earnings) has a long time horizon (i.e., changes, say, quarterly), then the corresponding risk factors are not relevant for shorter horizon strategies (e.g., those involving overnight returns),\footnote{\, Conversely, value-based longer horizon strategies would not benefit from any risk factors based on ``micro'' quantities with, say, millisecond horizons. E.g., statistical arbitrage strategies have high turnover as they attempt to capture intraday mean-reversion effects due to market over-/under-reactions to news events, etc. Value based strategies have very low turnover given that periods of extreme mispricings seldom occur (e.g., '87 Crash, '08 Meltdown).} unless there is a linear {\em extrapolating} argument that reasonably relates such longer term quantities to their shorter term counterparts (as, e.g., in the case of liquidity). More technically, suppose we have $K$ factors we know add value. How do we determine if a new, $(K+1)$-th, factor adds value?\footnote{\, In the next section we discuss why no-value-adding factors can increase trading costs.} Here is a simple method.

{}Thus, suppose we have $N$ stocks and we have FLM $\Omega_{iA}$, $i=1,\dots,N$, $A=1,\dots,K$. Let $\Omega^\prime_{iA^\prime}$, $i=1,\dots,N$, $A^\prime=1,\dots,K^\prime$, $K^\prime\equiv K+1$ be new FLM once we add a new, $(K+1)$-th, risk factor. (So we have $\left.\Omega^\prime_{iA^\prime}\right|_{A^\prime = A} = \Omega_{iA}$, $A=1,\dots,K$.) Let $R_i$ be the returns used in our strategy, i.e., these returns have the time horizon relevant to our strategy.\footnote{\, E.g., $R_i$ are overnight returns, we obtain alphas from these returns by regressing them (possibly, with some weights) over some FLM, and then we trade on these alphas right after the open.} We can run two regressions (without intercept -- unless it is included in $\Omega_{iA}$), first $R_i$ over $\Omega_{iA}$, and second $R_{i}$ over $\Omega^\prime_{i A^\prime}$. In R notations:
\begin{eqnarray}\label{rreg1}
 &&R \sim -1 + \Omega\\
 &&R \sim -1 + \Omega^\prime\label{rreg2}
\end{eqnarray}
In actuality, $R_i$, $\Omega_{iA}$ and $\Omega^\prime_{i A^\prime}$ are time series: $R_i(t_s) \equiv R_{si}$, $\Omega_{iA}(t_s) \equiv \Omega_{siA}$, $\Omega^\prime_{i A^\prime}(t_s) \equiv \Omega^\prime_{s i A^\prime}$, $s=0,1,\dots,M$. We can run the above two regressions for each value of $s$ and look at, e.g., two time-series vectors of the regression F-statistic to assess if the new risk factor improves the overall $F$-statistic.\footnote{\, To improve statistical significance, outliers can be removed (or smoothed, e.g., Winsorized).} Alternatively, we can pull the $(M+1)\times N$ matrix $R_{si}$ into a vector ${\widehat R}_\sigma$ of length $(M+1)N$ (i.e., treat the index pair $(s,i)$ as a single index $\sigma$), and do the same with FLM: $\Omega_{s i A}\equiv {\widehat \Omega}_{\sigma A}$, $\Omega^\prime_{s i A^\prime}\equiv {\widehat \Omega}^\prime_{\sigma A^\prime}$. We can now run two regressions
\begin{eqnarray}\label{reg1}
 &&{\widehat R} \sim -1 + {\widehat \Omega}\\
 &&{\widehat R} \sim -1 + {\widehat \Omega}^\prime\label{reg2}
\end{eqnarray}
and compare the F-statistic.\footnote{\, When assessing F-statistic, it needs to be taken into account that we have $(K+1)$ {\em vs.} $K$ factors, as is a possible change in the number of observations per factor due to any NAs.} If $K$ is not large, it is also informative to compare the t-values of the regression coefficients and assess the effect of the new factor.

{}For illustrative purposes, we ran such regressions for overnight returns $R_i \equiv \log(P_i(t_{\rm{\scriptstyle{open}}})/P_i(t_{\rm{\scriptstyle{close}}}))$, where the open $P_i(t_{\rm{\scriptstyle{open}}})$ and the previous close $P_i(t_{\rm{\scriptstyle{close}}})$ prices are adjusted for splits and dividends. In the case of, say, book value, as a benchmark it suffices to consider a $K=1$ model, where the sole risk factor is the intercept. Then we add the second risk factor, which is (log of) book (or tangible book, price-to-book, etc.),\footnote{\, We used fundamental data from stockpup.com (accessed 07/28/2014) and pricing data from finance.yahoo.com (accessed 07/29/2014) from 06/18/2009 through 06/20/2014 for a universe of 493 stocks, essentially from S\&P500. Negative (tangible) book values were omitted for the entire backtesting period.} so $K^\prime = 2$. The regression F-statistic and t-values are given in Table \ref{table1} (for regressions (\ref{reg1}) and (\ref{reg2})), which shows that the second regression (\ref{reg2}) involving (tangible) book value does not have improved statistic over the intercept-only regression. The 1-factor regressions other than the intercept-only regression can be thought of as regressions over ``betas''. The $\log(\mbox{Prc/Book})$ case (see Table \ref{table1}) is the closest to the intercept-only case because the regression $\mbox{Prc} \sim -1 + \mbox{Book}$ has F-statistic 56,230, and the t-value 237.1, i.e., price and book value are highly correlated. As to the 2-factor regressions, (T)Book does not improve the statistic. It is log(Prc) that makes impact, precisely because prices change daily.

{}We also ran the (\ref{reg1}) and (\ref{reg2}) regressions with a $K=10$ model as a benchmark, where the risk factors are 10 BICS sectors\footnote{\, Stocks rarely jump industries let alone sectors, so sector assignments are robust against time.} (so $K^\prime = 11$). The results are given in Table \ref{table2}, which shows that book value does not improve regression statistic. As above, it is log(Prc) that provides improvement. We also ran the (\ref{rreg1}) and (\ref{rreg2}) regressions separately for each date (i.e., without pulling the index pair $(s,i)$ into a single index $\sigma$ -- see above) with the same $K=10$ benchmark. The results are given in Table \ref{table2.5} and agree with those in Table \ref{table2}. Log(Prc), not Book, has impact.

{}Finally, for the (\ref{rreg1}) and (\ref{rreg2}) regressions we computed the t-statistic of actual risk factor time series {\em a la} Fama and MacBeth \cite{Fama.1973}, both for the $K=1$ (intercept only) and $K=10$ (BICS sectors) benchmark factor models. The results are given in Table \ref{table2.6} and Table \ref{table2.7} and agree with those in Table \ref{table1}, Table \ref{table2} and Table \ref{table2.5}.

\section{Pitfalls of Standardized Risk Models}\label{sec4}

\subsection{Industry Risk Factors}

{}Suppose we have an industry classification. For our discussion below it will make no difference whether FLM $\Omega^{\rm{\scriptstyle{ind}}}_{iA}$, $A=1,\dots,K_{\rm{\scriptstyle{ind}}}$ is binary or some conglomerates are allowed, so for simplicity we will assume it to be binary (see Subsection \ref{ind.class}):
\begin{equation}
 \Omega^{\rm{\scriptstyle{ind}}}_{iA} = \delta_{G(i), A}
\end{equation}
For definiteness, let us fix the names of the industry tree levels as ``sectors $\rightarrow$ sub-sectors $\rightarrow$ industries'', so ``industries'' correspond to the most detailed level. The number of the industries $K_{\rm{\scriptstyle{ind}}}$ depends on the universe. Different universes $U_1$ and $U_2$ can have vastly different industries to which the corresponding stocks belong.

{}In SRM a large number of stocks (e.g., several thousand for the U.S. models) are ``squeezed'' into a relatively modest number $K^*_{\rm{\scriptstyle{ind}}}$ of standardized industries, which can be substantially smaller than the number of true industries $K_{\rm{\scriptstyle{ind}}}$ for a typical quantitative trading portfolio universe of, say, 1,000-2,500 names. So, standardized industries typically lose granularity, which determines how well RM helps hedge a portfolio against industry risk. For illustrative purposes, let us look at the number of true industries for top-by-market-cap portfolios in BICS.\footnote{\, BICS naming convention is ``sectors $\rightarrow$ industries $\rightarrow$ sub-industries'', so our ``industries'' correspond to BICS ``sub-industries'', and our ``sub-sectors'' correspond to BICS ``industries''.} We require at least 10 stocks in each industry. Small industries are pruned to the sub-sector level and, if need be, to the sector level. Any leftover small industries can be merged into larger industries. The result is given in Table \ref{table3}. The numbers of true industries for top 1,500+ universes are sizably higher than those of typical standardized industries.

\subsection{Empty Standardized Industries}

{}Furthermore, for any given universe $U$, even if $U$ is, say, 1,000-2,500 names, we can have (almost) empty standardized industries -- e.g., a portfolio that does not trade any stocks from a given (sub-)sector. Empty standardized industries would have been omitted had we built RM based on the custom universe $U$. In SRM we have no such option, so we must keep empty industries. Why is this so bad?

{}Style factors are not important for our discussion here, so let us consider RM with only industry factors. Let these be $K$ standardized binary industries. Let the risk model universe be $U_{SRM}$, and let our universe be $U \subset U_{SRM}$. Let FLM be $\Omega_{iA}$, $i\in \{1,\dots,N\}\equiv U$. Let some industries be empty with $N_A = 0$, where
\begin{equation}
 N_A\equiv \sum_{i=1}^N \Omega_{iA} = \sum_{i=1}^N \delta_{G(i),A}
\end{equation}
Such industries must be omitted from regressions, if this is how we use RM.

{}On the other hand, suppose we are doing optimization, and we cannot omit such empty industries. Let $J\equiv\{A| N_A > 0\}$ and $J^\prime\equiv \{A| N_A = 0\}$. We have
\begin{equation}\label{empty.ind}
 \Gamma_{ij} = \xi_i^2~\delta_{ij} + \sum_{A,B\in J} \Omega_{iA}~\Omega_{jB}~\Phi_{AB} = \xi_i^2~\delta_{ij} + \Phi_{G(i), G(j)}
\end{equation}
The number of risk factors in this model is ${\widehat K} = |J| < K$, yet FCM $\Phi_{AB}$ is computed based on $K$ factors. So, from the viewpoint of the universe $U$, there are $|J^\prime| = K - {\widehat K}$ ``hidden'' factors, to which $U$ has no exposure, yet they affect the covariance matrix $\Gamma$. This does not bode well with the RM premise that all covariances are explained by a combination of: i) some fixed risk factors, exposure to which for a given universe of stocks is well-defined; and ii) ISR, which describes all uncertainty not described by the risk factors. Based on this premise, the correct way of modeling risk for our universe $U$ would be to assume that we have ${\widehat K}$ risk factors and compute FCM ${\widehat \Phi}_{AB}$, $A,B\in J$ for these factors along with the corresponding ISR ${\widehat \xi}_i$, i.e., to have
\begin{equation}\label{fixed.ind}
 {\widehat \Gamma}_{ij} \equiv {\widehat \xi}_i^2~\delta_{ij} + \sum_{A,B\in J} \Omega_{iA}~\Omega_{jB}~{\widehat \Phi}_{AB} = {\widehat \xi}_i^2~\delta_{ij} + {\widehat \Phi}_{G(i), G(j)}
\end{equation}
At first it might appear that (\ref{empty.ind}) and (\ref{fixed.ind}) are identical -- if $\Phi_{AB}$ and ${\widehat \Phi}_{AB}$ are computed as SCM of the corresponding risk factors, then we should have ${\widehat \Phi}_{AB} = \Phi_{AB}$, $A,B\in J$. However, as mentioned in Subsection \ref{fac.cov}, in real life FCM is {\em not} computed as SCM, because ISR with such a computation are typically ill-defined. Consequently, ${\widehat \Phi}_{AB} \neq \Phi_{AB}$, $A,B\in J$ and ${\widehat \xi}_i\neq \xi_i$. Because of this interdependency of FCM and ISR, it is more desirable to compute ${\widehat \Phi}_{AB}$ and ${\widehat \xi}_i$ directly based on the universe $U$ without any empty industries -- the latter only bring more noise into the computation. Any uncertainty not described by the relevant risk factors should be modeled via ISR, not some ``hidden'' risk factors.

{}That empty industries, to which the universe $U$ has no exposure, add noise can be seen in the optimization context -- they contribute to the desired dollar holdings (\ref{d.h}) via (\ref{Q.tilde}).\footnote{\, FCM and ISR must be recomputed based on non-empty industries to remove this contribution.} The effect is that the desired dollar holdings are approximately neutralized against these ``hidden'' industries (see (\ref{r.n.1}) and (\ref{r.n.2})).\footnote{\, These ``hidden'' industries might be correlated with the non-empty industries. However, such correlations should not be high -- if they are high, then the industry classification is too granular (or deficient) and must be pruned (or replaced with a more precise industry classification). Including redundant noise-generating industries in RM is certainly not the right way to handle such cases.} Typically, this generates additional noise (``twitch'') trades, which on paper may appear harmless,\footnote{\, On paper such noise trades typically have little effect on the simulated P\&L, but can reduce the Sharpe ratio -- if the approximate neutrality constraints effected by these empty industries do not add value, then the portfolio is suboptimal, i.e., it does not maximize the Sharpe ratio.} but in real life can increase trading costs and reduce profitability, rendering empty industries undesirable. The same argument applies to irrelevant style factors (e.g., value/growth in short horizon strategies; see Section \ref{sub3.1}), rendering them harmful.\footnote{\, This brings us to another point we made in Introduction: in SRM typically FCM is computed based on some universe $U_*$, a fraction of $U_{SRM}$. For the same reasons as above, it is preferable to compute FCM based on the trading universe $U$ as $U_*$ may not have substantial overlap with $U$.}

\subsection{Style Risk Factors}

{}Some -- albeit not necessarily all\footnote{\, E.g., style factors with discrete values are not normalized. An example is a binary style factor in some SRM indicating whether the stock belongs to the universe $U_*$ defined above.} -- style factors are normalized. One way of normalizing a style factor labeled by $A\in\{1,\dots,K\}$ is by conforming the values of the A-th column in FLM $\Omega_{iA}$ to a normal distribution with, e.g.,  mean 0 and standard deviation equal to the standard deviation (or MAD) of the original, unnormalized column. This is done in the R code in Appendix \ref{appA} for momentum, size and liquidity. If the distribution is expected to be log-normal, then one normalizes log of the column and re-exponentiates. This is done for intraday volatility in the R code in Appendix \ref{appA}.

{}Such normalizations of style factors are typically done across the entire coverage universe $U_{SRM}$. Suppose we wish to use SRM for our universe $U$, which is a fraction of $U_{SRM}$. Then the values of the corresponding columns in $\Omega_{iA}$ with $i\in U$ are no longer normalized. For a random subset $U\subset U_{SRM}$ one may expect that they are still approximately normalized if the number of stocks in $U$ is large. However, typically there is nothing random about a real-life trading universe $U$, which is carefully selected based on certain requirements on market cap, liquidity, volatility, price and/or other relevant quantities, which can be rather skewed, so the truncated to $U$ style factor columns in $\Omega_{iA}$ are no longer even approximately normalized. Why does this matter? If FCM is computed based on normalized style factors (be it using $U_{SRM}$ or $U_*$), then this FCM is not the same (or even necessarily close to) FCM one would obtain based on style factors normalized based on $U$. This also affects ISR. Therefore, for the same reasons as in the previous subsection, it is more desirable to compute FLM, FCM and ISR based on the custom universe $U$.

\section{Concluding Remarks}\label{sec5}

{}Above we discussed points (1)-(4) raised in Introduction in relation to using SRM and when and why CRM are warranted. Let us conclude by briefly touching upon points (5)-(7) mentioned in Introduction, which relate to other aspects.

{}Statistical Arbitrage (StatArb) ``refers to highly technical short-term mean-reversion strategies involving large numbers of securities (hundreds to thousands, depending on the amount of risk capital), very short holding periods (measured in days to seconds), and substantial computational, trading, and information technology (IT) infrastructure'' \cite{Lo.2010}. A quantitative framework for this ``mean-reversion'' was recently discussed in \cite{Kakushadze}. Schematically, one can think about this mean-reversion as follows. Pick some returns. Pick RM. Then: i) either regress your returns over FLM (or a subset of its columns) with some weights, or ii) do optimization using RM (possibly with some constraints). One can hang various bells and whistles onto a strategy constructed this way, which do contribute into differentiating models. Nonetheless, the choice of RM is a major factor (along with the choice of returns) in what the desired holdings and trades will look like. As discussed in Section \ref{risk.taking}, in optimization desired holdings are approximately neutral w.r.t the risk factors, while in regression they are exactly neutral. So RM, in its uses discussed above, factors away (completely or approximately) the risk exposure in the returns w.r.t. the risk factors. Therefore, using the same RM for two sets of apparently different returns can substantially reduce the difference between the two -- when it comes to the resulting desired holdings -- making the two strategies more correlated.

{}Therefore, to diversify strategies it is not only important to diversify the returns, but also to diversify RM. This is especially useful in the ``herd effect'' situations, where many market participants -- for whatever (temporary) underlying reason ({\em cf.} Aug'07 Quant Meltdown) -- are ``compelled'' to do the same thing. When it comes to unpleasant situations such as potential book liquidation, being less correlated with others can make a difference between liquidating (and incurring huge transaction costs) or weathering the storm. Custom RM in such cases can make a difference.

{}We can take this further by noting that, even with the same returns, two substantially different RM, call them RM-A and RM-B, can produce P\&L streams which are not that highly correlated. In this case, instead of running one strategy with the returns, say, optimized using only RM-A, it makes more sense to run two strategies -- seamlessly, with trades crossed internally at the level of desired holdings, i.e., the two strategies are combined with some weights $w_A$ and $w_B$ (see below) -- where the first strategy, call it Str-A, is optimized using RM-A, and the second strategy, call it Str-B, is optimized using RM-B. If both strategies have positive returns and are not too highly correlated, then even if Str-B has worse return than Str-A, it still makes sense to combine them with some weights. In the zeroth approximation the weights $w_A$ and $w_B$ can be obtained, e.g., by requiring that the Sharpe ratio of the resulting combined strategy be maximized. However, in real life -- so long as the Sharpe ratio is acceptably high -- typically it is the P\&L that matters. In this regard, combining the two strategies can increase the P\&L as the capacity bound of the combined strategy is higher (than those of Str-A and Str-B) -- the Str-A and Str-B trades are not highly correlated and by combining them one reduces turnover, thereby decreasing market impact and increasing capacity, and also decreasing costs of trading. Turnover reduction and increasing capacity are two important incentives for diversifying and using CRM.

{}Finally, let us mention that CRM provide additional evident benefits listed in point (7) in Introduction. Also, for shorter-term applications (which is precisely when CRM are warranted), which do not require long lookbacks or nontrivial fundamental data (such as earnings going back many years), the data required for building a CRM is typically already available to the portfolio manager. All in all, it boils down to computing FCM and ISR in a self-consistent fashion.

\appendix

\section{R Code for Some Style Risk Factors}\label{appA}

{}Below we give R code for 4 style factors (function {\tt{calc.add.fac()}}): momentum, liquidity, size and intraday volatility. All uninformative dependencies have been omitted. The code is unaltered otherwise. (Functions {\tt{normalize()}}, {\tt{calc.sr()}}, {\tt{calc.eff.mad()}}, {\tt{calc.ret.mv()}} and {\tt{calc.ret.mv.clean()}} are auxiliary.) Below: {\tt{is.adr}} is a binary $N$-vector ($N$ is the number of stocks); {\tt{hist.prc, hist.vol, hist.high, hist.low, hist.cap}} are $d \times N$ matrices ($d$ is the number of trading dates in the historical data) -- historical closing price, daily volume, daily high, daily low (all adjusted -- this is necessary for momentum, albeit not for liquidity or intraday volatility), and market cap, respectively; {\tt{dates}} is the vector of the last 252 trading dates. ADRs are normalized independently. {\tt{days}} is the lookback (e.g., 252 trading days), while {\tt{back}} is for out-of-sample backtesting (and is 0 for current-date calculations); both {\tt{days}} and {\tt{back}} are passed into {\tt{calc.add.fac()}} via the omitted arguments.\\
\\
{\tt{\small
\noindent normalize <- function(x, center = mean(x), sdev = sd(x))\{\\
\indent	return(qnorm(ppoints(x)[sort.list(sort.list(x), method = ''radix'')],\\
\indent\indent center, sdev))\\
\}\\
\noindent calc.sr <- function(tv)\{\\
\indent	sr <- sqrt(tv)\\
\indent	sr <- log(sr)\\
\indent	sr <- normalize(sr, median(sr), mad(sr))\\
\indent	sr <- exp(sr)\\
\indent	return(sr)\\
\}\\
\noindent calc.eff.mad <- function(ret)\{\\
\indent eff.mad <- (5 * outer(apply(ret, 1,\\
\indent \indent mad, na.rm = T), apply(ret,\\
\indent 2, mad, na.rm = T), pmax))\\
\indent return(eff.mad)\\
\}\\
\noindent calc.ret.mv <- function(prc, back, days, d.r)\{\\
\indent last <- nrow(prc) - back\\
\indent first <- last - days\\
\indent today <- prc[last:first,  ]\\
\indent yest <- 0\\
\indent for(i in 1:d.r)\\
\indent \indent yest <- yest + prc[(last - i):(first - i), ]\\
\indent yest <- yest / d.r\\
\indent ret <- today/yest - 1\\
\indent dimnames(ret) <- dimnames(prc[last:first,  ])\\
\indent return(ret)\\
\}\\
\noindent calc.ret.mv.clean <- function(prc, back, days, d.r)\{\\
\indent ret <- calc.ret.mv(prc, back, days, d.r)\\
\indent eff.mad <- calc.eff.mad(ret)\\
\indent bad <- abs(ret - apply(ret, 1, median)) > eff.mad\\
\indent ret[bad] <- NA\\
\indent avg.ret <- matrix(rowMeans(ret, na.rm = T),\\
\indent\indent nrow = nrow(ret), ncol = ncol(ret))\\
\indent ret[bad] <- avg.ret[bad]\\
\indent ret <- ret - avg.ret\\
\indent return(ret)\\
\}\\
calc.add.fac <- function(...)\{\\
\indent	\#\#\# MOMENTUM MOVING AVERAGE LENGTHS
\indent	d.r <- 5\\
\\
\indent	\#\#\# ADDV MOVING AVERAGE LENGTHS\\	
\indent	d.addv <- 20\\
\\
\indent \#\#\# MOMENTUM FACTOR\\
\indent \#\#\# BASED ON AVERAGE 5-DAY RETURNS (OUTLIERS REMOVED)\\
\indent ret.mom <- calc.ret.mv.clean(hist.prc, back, days, d.r)\\
\indent mom <- apply(ret.mom, 2, mean)\\
\indent mom <- normalize(mom, 0, mad(mom))\\
\\
\indent	\#\#\# AVERAGE DAILY DOLLAR VOLUME (ADDV) FACTOR\\
\indent	\#\#\# BASED ON LAST 20 DAYS\\
\indent	\#\#\# ADRS ARE NORMALIZED ACCORDING TO NON-ADR DISTRIBUTION\\
\indent	not.adr <- !is.adr\\
\indent	addv <- hist.prc[dates, ] * hist.vol[dates, ]\\
\indent	addv <- addv[1:d.addv, ]\\
\indent	addv[addv == 0] <- NA\\
\indent	addv <- colMeans(log(addv), na.rm = T)\\
\indent	addv[is.adr] <- normalize(addv[is.adr], 0, mad(addv[not.adr]))\\
\indent	addv <- normalize(addv, 0, mad(addv[not.adr]))\\
{}\\
\indent	\#\#\# MARKET CAP FACTOR\\
\indent	\#\#\# BASED ON 252 DAYS\\
\indent	\#\#\# ADRS ARE NORMALIZED ACCORDING TO NON-ADR DISTRIBUTION\\
\indent	mkt.cap <- hist.cap[dates, ]\\
\indent	mkt.cap[mkt.cap == 0] <- NA\\
\indent	mkt.cap <- colMeans(log(mkt.cap), na.rm = T)\\
\indent	mkt.cap[is.adr] <- normalize(mkt.cap[is.adr], 0, mad(mkt.cap[not.adr]))\\
\indent	mkt.cap <- normalize(mkt.cap, 0, mad(mkt.cap[not.adr]))\\
{}\\
\indent	\#\#\# INTRADAY VOLATILITY FACTOR\\
\indent	\#\#\# BASED ON 252 DAYS\\
\indent	hist.low <- hist.low[dates, ]\\
\indent	hist.high <- hist.high[dates, ]\\
\indent	hist.prc <- hist.prc[dates, ]\\
\indent	high.low <- abs(hist.high - hist.low) / hist.prc\\
\indent	high.low <- calc.sr(colMeans(high.low\^{}2))\\
\}
}}

\section{C Code for Symmetric Matrix Inversion}\label{appB}

{}The vector {\tt{a[]}} is a symmetric $n\times n$ matrix $A_{ij}$ to be inverted pulled into a vector such that $A_{ij}$, $i,j = 0,1,\dots,n-1$ is given by {\tt{a[i + n * j]}}. This algorithm utilizes the fact that the matrix is symmetric thereby reducing the number of operations (compared with the Gauss-Jordan method for a general matrix).\\
\\
{\tt{\small
\noindent static void InvSymMat(double *a, int n)\{\\
\indent	int i, j, k;\\
\indent	double sum;\\

\indent	for( i = 0; i < n; i++ )\\
\indent	\indent	for( j = i; j < n; j++ )\\
\indent\indent		\{\\
\indent\indent\indent           sum = a[i + n * j];\\
\\
\indent\indent\indent			for( k = i - 1; k >= 0; k-- )\\
\indent\indent\indent\indent				sum -= a[i + n * k] * a[j + n * k];\\
\\
\indent\indent\indent			a[j + n * i] = \\
\indent\indent\indent\indent        ( j == i ) ? 1 / sqrt(sum) : sum * a[i * (n + 1)];\\
\indent\indent		\}\\
\\
\indent	for( i = 0; i < n; i++ )\\
\indent\indent		for( j = i + 1; j < n; j++ )\\
\indent\indent		\{\\
\indent\indent\indent			sum = 0;\\
\\
\indent\indent\indent			for( k = i; k < j; k++ )\\
\indent\indent\indent\indent				sum -= a[j + n * k] * a[k + n * i];\\
\\
\indent\indent\indent			a[j + i * n] = sum * a[j * (n + 1)];\\
\indent\indent		\}\\
\\
\indent	for( i = 0; i < n; i++ )\\
\indent\indent		for( j = i; j < n; j++ )\\
\indent\indent		\{\\
\indent\indent\indent			sum = 0;\\
\indent\indent\indent			for( k = j; k < n; k++ )\\
\indent\indent\indent\indent				sum += a[k + n * i] * a[k + n * j];\\
\\
\indent\indent\indent			a[i + n * j] = a[j + n * i] = sum;\\
\indent\indent		\}\\
\}
}}

\section{DISCLAIMERS}\label{appC}

{}Wherever the context so requires, the masculine gender includes the feminine and/or neuter, and the singular form includes the plural and {\em vice versa}. The author of this paper (``Author'') and his affiliates including without limitation Quantigic$^\circledR$ Solutions LLC (``Author's Affiliates'' or ``his Affiliates'') make no implied or express warranties or any other representations whatsoever, including without limitation implied warranties of merchantability and fitness for a particular purpose, in connection with or with regard to the content of this paper including without limitation any code or algorithms contained herein (``Content'').

{}The reader may use the Content solely at his/her/its own risk and the reader shall have no claims whatsoever against the Author or his Affiliates and the Author and his Affiliates shall have no liability whatsoever to the reader or any third party whatsoever for any loss, expense, opportunity cost, damages or any other adverse effects whatsoever relating to or arising from the use of the Content by the reader including without any limitation whatsoever: any direct, indirect, incidental, special, consequential or any other damages incurred by the reader, however caused and under any theory of liability; any loss of profit (whether incurred directly or indirectly), any loss of goodwill or reputation, any loss of data suffered, cost of procurement of substitute goods or services, or any other tangible or intangible loss; any reliance placed by the reader on the completeness, accuracy or existence of the Content or any other effect of using the Content; and any and all other adversities or negative effects the reader might encounter in using the Content irrespective of whether the Author or his Affiliates is or are or should have been aware of such adversities or negative effects.

{}The R code included in Appendix \ref{appA} hereof is part of the copyrighted R code for Quantigic$^\circledR$ Risk Model$^{\rm{\scriptstyle{TM}}}$ and is provided herein with the express permission of Quantigic$^\circledR$ Solutions LLC. The copyright owner retains all rights, title and interest in and to its copyrighted source code included in Appendix \ref{appA} and Appendix \ref{appB} hereof and any and all copyrights therefor.

{}The Content is not intended, and should not be construed, as an investment, legal, tax or any other such advice, and in no way represents views of Quantigic$^\circledR$ Solutions LLC, the website \underline{www.quantigic.com} or any of their other affiliates.

\newpage
\begin{table}[ht]
\caption{Results for regressions (\ref{reg1}) and (\ref{reg2}) with the intercept-only 1-factor model as the benchmark. Int = intercept; (T)Book = (tangible) book; Prc = adjusted previous close; RPrc = raw (unadjusted) previous close. Next to Int+log(Prc) we also give Int+log(RPrc) results. We do this because adjusting the previous close introduces a bias of anticipating future splits and/or dividends. However, as can be seen from the Int+log(RPrc) row, this bias is relatively mild and does not affect our conclusions.} 
\begin{tabular}{l l l l} 
\\
\hline\hline 
Regression/Statistic & F-statistic & Intercept & Second Coefficient \\
& & t-value & t-value\\ [0.5ex] 
\hline 
Int only & 737.7 & 27.16 & --- \\ 
Book only & 237.2 & --- & 15.40\\
TBook only & 191.2 & --- & 13.83\\
Prc only & 1.34 & --- & 1.16\\
Prc/Book only & 12.5 & --- & 3.54\\
Prc/Tbook only & 3.84 & --- & 1.96\\
log(Book) only & 707.5 & --- & 26.60\\
log(TBook) only & 583.7 & --- & 24.70\\
log(Prc) only & 526.0 & --- & 22.94\\
log(Prc/Book) only & 739.1 & --- & -27.19\\
log(Prc/Tbook) only & 608.7 & --- & -24.67\\
Int+Book & 362.5 & 22.08 & 4.10 \\
Int+TBook & 297.6 & 20.10 & 4.56 \\
Int+(Prc/Book) & 354.3 & 26.38 & -0.66 \\
Int+(Prc/TBook) & 287.2 & 23.89 & 0.15\\
Int+Prc & 368.9 & 27.14 & -0.24\\
Int+log(Book) & 354.2 & 0.98 & 0.53 \\
Int+log(TBook) & 294.1 & -2.11 & 3.70 \\
Int+log(Prc) & 473.9 & 20.53 & -14.48 \\
Int+log(RPrc) & 468.7 & 20.18 & -14.12 \\
Int+log(Prc/Book) & 394.0 & -6.99 & -8.93 \\
Int+log(Prc/TBook) & 329.9 & -7.14 & -9.23 \\ [1ex] 
\hline 
\end{tabular}
\label{table1} 
\end{table}

\begin{table}[ht]
\caption{Results for regressions (\ref{reg1}) and (\ref{reg2}) with the BICS-sector 10-factor model as the benchmark. S = 10 BICS sectors labeled by S1(30), S2(63), S3(45), S4(30), S5(91), S6(75), S7(42), S8(48), S9(41) and S10(28) (the parentheticals show the number of tickers in each sector); X = the 11th factor (P, B, P/B, log(P), log(B) and log(P/B)); P = adjusted previous close; B = book; F = F-statistic; t = t-value. {\em E.g.}, in the ``Reg:" line ``S+(P/B)" means that the returns $R$ are regressed over FLM  $\Omega$ containing 11 columns corresponding to the 10 sectors S1 through S10 plus the 11th factor X, which is (P/B) in this case. In the S+log(P) column we also give the values when P is taken to be the raw (unadjusted) previous close. We do this because adjusting the previous close introduces a bias of anticipating future splits and/or dividends. However, as can be seen from the S+log(P) column, this bias is relatively mild and does not affect our conclusions.} 
\begin{tabular}{l l l l l l l l} 
\\
\hline\hline 
Reg: & S & S+P & S+B & S+(P/B) & S+log(P) & S+log(B) & S+log(P/B)\\[0.5ex] 
\hline 
F & 80.6 & 73.3 & 72.4 & 71.3 & 92.6/91.7 & 71.3 & 77.8\\ 
t:S1 & 6.40 & 6.40 & 6.16 & 6.40 & 15.07/14.80 & 1.56 & -6.42\\
t:S2 & 6.67 & 6.67 & 5.94 & 6.50 & 15.74/15.44 & 1.19 & -7.08\\
t:S3 & 5.57 & 5.57 & 5.02 & 5.60 & 15.08/14.76 & 1.56 & -7.04\\
t:S4 & 5.40 & 5.40 & 5.08 & 5.41 & 14.13/13.87 & 1.32 & -6.79\\
t:S5 & 13.31 & 13.28 & 11.12 & 13.36 & 19.26/18.93 & 1.80 & -6.38\\
t:S6 & 13.13 & 13.13 & 12.26 & 12.50 & 19.29/18.99 & 1.98 & -6.09\\
t:S7 & 4.97 & 4.97 & 3.25 & 3.74 & 14.40/14.15 & 0.89 & -7.37\\
t:S8 & 6.85 & 6.85 & 6.42 & 6.88 & 15.88/15.58 & 1.35 & -6.80\\
t:S9 & 12.83 & 12.84 & 11.90 & 12.92 & 19.40/19.12 & 2.49 & -5.39\\
t:S10 & 8.63 & 8.63 & 8.21 & 8.87 & 16.17/15.92 & 2.19 & -5.69\\
t:X & --- & -0.42 & 3.42 & -0.45 & -14.57/-14.21 & -0.20 & -8.47\\ [1ex] 
\hline 
\end{tabular}
\label{table2} 
\end{table}

\begin{table}[ht]
\caption{Results for regressions (\ref{rreg1}) and (\ref{rreg2}) with the BICS-sector 10-factor model as the benchmark. The notations are the same as in Table \ref{table2}, except that F = median F-statistic, and t = median t-value, where F-statistic and t-values are computed based on regressions (\ref{rreg1}) and (\ref{rreg2}) for each date, and the median is computed serially over all dates. The meaning of double entries in the S+log(P) column is the same as in Table \ref{table2}.} 
\begin{tabular}{l l l l l l l l} 
\\
\hline\hline 
Reg: & S & S+P & S+B & S+(P/B) & S+log(P) & S+log(B) & S+log(P/B)\\[0.5ex] 
\hline 
F & 13.5 & 12.2 & 12.0 & 12.0 & 12.3/12.3 & 12.1 & 12.2\\ 
t:S1 & 0.40 & 0.40 & 0.40 & 0.43 & 0.67/0.68 & 0.11 & -0.18\\
t:S2 & 0.61 & 0.61 & 0.62 & 0.63 & 0.85/0.83 & 0.11 & -0.16\\
t:S3 & 0.34 & 0.33 & 0.32 & 0.32 & 0.69/0.63 & 0.11 & -0.20\\
t:S4 & 0.23 & 0.23 & 0.23 & 0.25 & 0.62/0.56 & 0.10 & -0.21\\
t:S5 & 0.91 & 0.91 & 0.76 & 0.88 & 0.90/0.85 & 0.15 & -0.19\\
t:S6 & 0.80 & 0.80 & 0.82 & 0.91 & 0.96/0.92 & 0.14 & -0.14\\
t:S7 & 0.39 & 0.39 & 0.24 & 0.25 & 0.70/0.64 & 0.10 & -0.20\\
t:S8 & 0.54 & 0.54 & 0.57 & 0.57 & 0.76/0.74 & 0.08 & -0.19\\
t:S9 & 0.76 & 0.76 & 0.77 & 0.80 & 1.00/0.98 & 0.18 & -0.12\\
t:S10 & 0.53 & 0.53 & 0.52 & 0.55 & 0.84/0.88 & 0.14 & -0.14\\
t:X & --- & -0.02 & 0.13 & -0.04 & -0.55/-0.52 & -0.03 & -0.30\\ [1ex] 
\hline 
\end{tabular}
\label{table2.5} 
\end{table}

\begin{table}[ht]
\caption{Results for regressions (\ref{rreg1}) and (\ref{rreg2}) with the intercept-only 1-factor model as the benchmark. The notations are the same as in Table \ref{table1}, except that the t-statistic here refers to the t-statistic of the corresponding risk factor time series {\em a la} Fama and MacBeth (1973). These t-statistic are annualized, {\em i.e.}, we compute the daily t-statistic and then multiply it by $\sqrt{252}$.} 
\begin{tabular}{l l l l} 
\\
\hline\hline 
Regression/Statistic & Intercept & Second Coefficient \\
& t-statistic & t-statistic\\ [0.5ex] 
\hline 
Int only & 0.90 & --- \\ 
Int+Book & 0.82 & 2.21 \\
Int+(Prc/Book) & 0.90 & -0.69 \\
Int+Prc & 0.90 & -0.42\\
Int+log(Book) & 0.23 & 0.32 \\
Int+log(Prc) & 1.90 & -3.50 \\
Int+log(RPrc) & 1.78 & -3.10 \\
Int+log(Prc/Book) & -2.15 & -2.90\\ [1ex] 
\hline 
\end{tabular}
\label{table2.6} 
\end{table}

\begin{table}[ht]
\caption{Results for regressions (\ref{rreg1}) and (\ref{rreg2}) with the BICS-sector 10-factor model as the benchmark. The notations are the same as in Table \ref{table2}, except that ``t:$\star$" refers to the annualized t-statistic of the corresponding risk factor ``$\star$" time-series {\em a la} Fama and MacBeth (1973), same as in Table \ref{table2.6}. The meaning of double entries in the S+log(P) column is the same as in Table \ref{table2}.} 
\begin{tabular}{l l l l l l l l} 
\\
\hline\hline 
Reg: & S & S+P & S+B & S+(P/B) & S+log(P) & S+log(B) & S+log(P/B)\\[0.5ex] 
\hline 
t:S1 & 0.76 & 0.76 & 0.74 & 0.78 & 1.76/1.64 & 0.39 & -2.02\\
t:S2 & 0.59 & 0.59 & 0.54 & 0.60 & 1.61/1.50 & 0.28 & -2.24\\
t:S3 & 0.76 & 0.76 & 0.68 & 0.77 & 2.07/1.91 & 0.30 & -2.42\\
t:S4 & 1.09 & 1.09 & 1.01 & 1.10 & 2.32/2.14 & 0.37 & -2.48\\
t:S5 & 0.88 & 0.88 & 0.77 & 0.88 & 1.75/1.65 & 0.45 & -2.02\\
t:S6 & 1.07 & 1.07 & 1.02 & 1.05 & 2.01/1.88 & 0.51 & -1.86\\
t:S7 & 0.83 & 0.83 & 0.56 & 0.66 & 2.14/1.99 & 0.22 & -2.66\\
t:S8 & 0.64 & 0.64 & 0.60 & 0.65 & 1.72/1.59 & 0.32 & -2.07\\
t:S9 & 1.09 & 1.09 & 1.02 & 1.09 & 1.87/1.77 & 0.61 & -1.51\\
t:S10 & 1.17 & 1.17 & 1.13 & 1.23 & 2.04/1.92 & 0.58 & -1.78\\
t:X & --- & -0.56 & 2.15 & -0.61 & -3.45/-3.05 & 0.02 & -3.12\\ [1ex] 
\hline 
\end{tabular}
\label{table2.7} 
\end{table}


\begin{table}[ht]
\caption{Number of BICS industries for portfolios of stocks in top X by market cap with at least 10 stocks in each industry as of August 19, 2014. Only U.S. listed common stocks and class shares are included (no OTCs, preferred shares, {\em etc.}).} 
\begin{tabular}{l l l l l l l l} 
\\
\hline\hline 
Top X by market cap & Number of industries\\[0.5ex] 
\hline 
1,000 & 55\\ 
1,500 & 75\\
2,000 & 94\\
2,500 & 107\\
3,000 & 122\\
3,500 & 125\\
4,000 & 128\\
4,500 & 130\\
5,000 & 133\\ [1ex] 
\hline 
\end{tabular}
\label{table3} 
\end{table}

\end{document}